# Análisis de un oscilador no lineal utilizando el método de Euler en una planilla de cálculo

# Using the Euler method in a spreadsheet to study a nonlinear oscillator

Álvaro Suárez[1], Fernando Tornaría[2]

[1]*Departamento de Física del Instituto de Profesores Artigas, Montevideo, Uruguay; alsua@outlook.com.*
[2]*CES-ANEP, Montevideo, Uruguay; ftornaria@uruguayeduca.edu.uy.*

***Resumen:*** Se analizan teórica y experimentalmente las oscilaciones unidimensionales no lineales de un imán que está unido a un resorte e interactúa con un grupo de imanes. Se resuelve numéricamente la ecuación de movimiento del sistema utilizando el método de Euler en una planilla de cálculo. Experimentalmente, la evolución temporal de la posición se obtiene mediante análisis de video. La correspondencia entre las predicciones del modelo y los datos experimentales deja en evidencia la potencialidad de las herramientas utilizadas. La sencillez de la propuesta presentada permite su aplicación en los primeros cursos universitarios de Física experimental.

***Abstract:*** The one-dimensional nonlinear oscillations of a magnet attached to a spring and interacting with a set of magnets are analyzed theoretically and experimentally. The equation of motion is solved numerically using the Euler method in a spreadsheet. The temporal evolution of the position is obtained using video analysis. The correspondence between the data and the predictions of the model shows the potential of the tools employed. The simplicity of the approach allows its application in the first university courses of experimental Physics.

**1. Introducción**

En general, en cursos básicos de Física, los sistemas oscilatorios -así como otros sistemas físicos- se estudian bajo aproximaciones lineales. De esta forma pueden ser modelados con sencillez y resueltos analíticamente en forma explícita. Esto permite ganar en simplicidad, pero se deja entrever al estudiante que los problemas reales siempre pueden ser resueltos en forma exacta. Cuando los problemas no pueden ser descritos adecuadamente a partir de ecuaciones diferenciales con solución analítica, entra en juego la resolución numérica de las mismas. Las herramientas matemáticas para este tipo de tratamiento, así como el análisis de situaciones con esas características, son en general pospuestas para cursos avanzados de Física.

En este artículo nos centramos en el estudio teórico y experimental de un sistema oscilatorio cuya ecuación diferencial del movimiento no es lineal. El mismo está conformado por una pesa colgada de un resorte, sobre la cual también actúan fuerzas de origen magnético debido a la presencia de imanes. Se realiza un abordaje plausible de ser aplicado en cursos universitarios básicos de Física Experimental. Para determinar las variables dinámicas y cinemáticas del sistema, se resuelve numéricamente la ecuación diferencial de movimiento, aplicándose el método de Euler [1]. La justificación del uso de este procedimiento se basa en que sus predicciones son acordes a los resultados experimentales y que es fácil de comprender por los estudiantes [2]. De esta manera los estudiantes pueden trabajar con sistemas dinámicamente más complejos, cuya evolución temporal no se conoce a priori. Por otro lado, la programación del método de Euler es relativamente sencilla, pudiendo ser realizada por cualquier estudiante con una planilla de cálculo.

En la siguiente sección de este trabajo se describe con detalle el dispositivo experimental. En la sección 3, se describe cómo se caracterizaron las fuerzas involucradas sobre el sistema. En la sección 4 se analizan los resultados obtenidos y en la última sección se presentan las principales conclusiones.

## 2. Procedimiento experimental

El dispositivo experimental consiste en una pesa cilíndrica[1] que cuelga de un resorte[2] cuyo extremo superior está fijo (ver figura 1). En la base de la pesa se ubicaron cinco pequeños imanes cilíndricos de neodimio[3] y bajo los mismos se colocó un disco de metal[4]. Tanto la pesa como el disco eran de aleaciones ferromagnéticas y podían oscilar solidarios al resorte. Debajo de este sistema se encontraba fijo un segundo disco (similar al anterior) en cuya base había dos imanes de neodimio con forma de barra.[5]

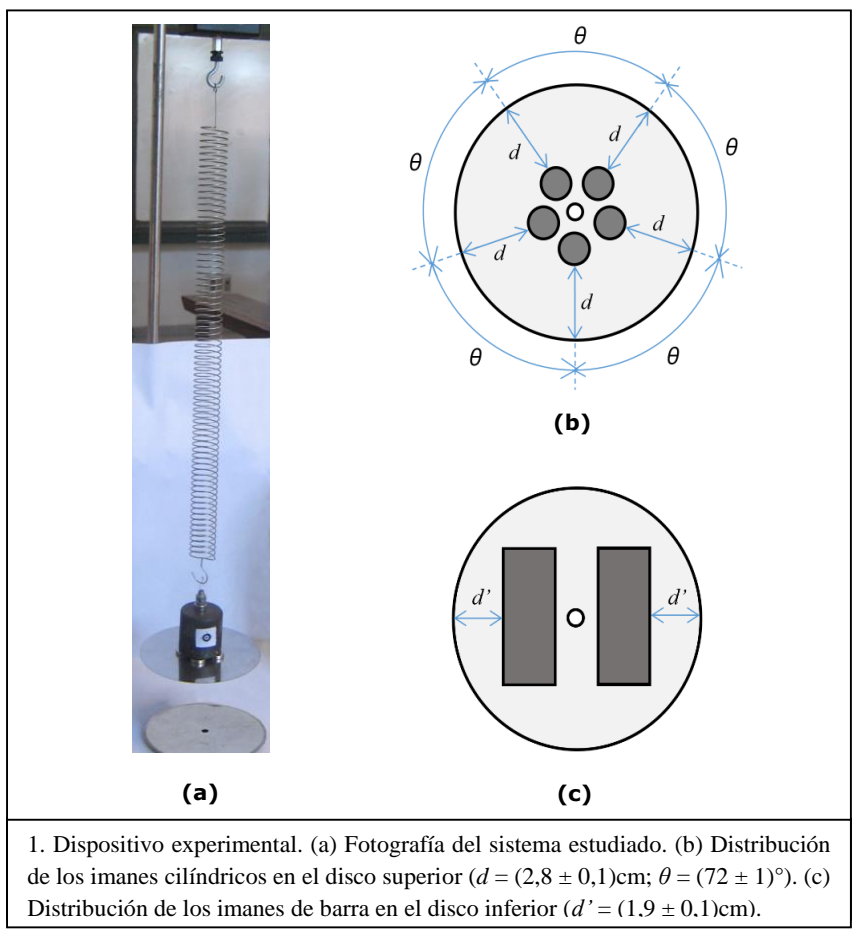

1. Dispositivo experimental. (a) Fotografía del sistema estudiado. (b) Distribución de los imanes cilíndricos en el disco superior ($d = (2,8 \pm 0,1)$cm; $\theta = (72 \pm 1)°$). (c) Distribución de los imanes de barra en el disco inferior ($d' = (1,9 \pm 0,1)$cm).

---

[1] Características de la pesa: altura = $(3,5 \pm 0,1)$cm; diámetro $(3,3 \pm 0,1)$cm; masa = $(213,7 \pm 0,1)$g.
[2] Masa del resorte: $(22,76 \pm 0,01)$g
[3] Característica de los imanes cilíndricos: diámetro = $(1,2 \pm 0,1)$cm; altura = $(0,30 \pm 0,01)$cm; masa = $(2,5 \pm 0,1)$g; intensidad del campo magnético a 1cm de distancia: $\sim 10^{-2}$ T.
[4] Característica del disco: diámetro = $(9,6 \pm 0,1)$cm; espesor = $(0,05 \pm 0,01)$cm; masa = $(31,7 \pm 0,1)$g; diámetro del orificio central: $(0,5 \pm 0,1)$cm.
[5] Característica de los imanes en forma de barra: largo = $(5,0 \pm 0,1)$cm; ancho = $(2,0 \pm 0,1)$cm; espesor = $(0,50 \pm 0,01)$cm; masa = $(37,2 \pm 0,1)$g; intensidad del campo magnético a 1cm de distancia: $\sim 10^{-1}$ T.

Para estudiar el movimiento de la pesa oscilando se registró el mismo con una cámara de video montada en un trípode.[6] La filmación obtenida fue analizada con el *software* Tracker [3-4]. Empleando este programa se determinó la posición de la pesa en cada uno de los fotogramas (ver figura 2). Para ello se empleó la función *autotracker*, que permite seleccionar una región en uno de los cuadros del video y realizar un seguimiento automático de la misma en el resto de la filmación. Para facilitar esta tarea se le colocó a la pesa un papel blanco con un dibujo de tres círculos concéntricos (de color blanco el de diámetro intermedio y de color negro los dos restantes). En el primer cuadro del video se seleccionó este diseño para su seguimiento automático con *autotracker*.

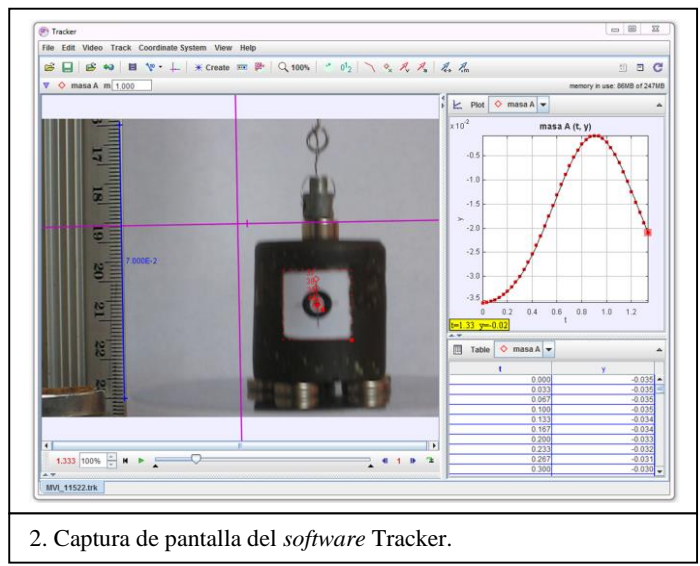

2. Captura de pantalla del *software* Tracker.

Como referencia de longitud se empleó una regla. Para minimizar los errores de paralaje, se prestó especial atención a que la regla y la trayectoria de la pesa estuvieran en un mismo plano. Y que a su vez dicho plano fuera paralelo al objetivo de la cámara. Con el fin de minimizar la distorsión introducida por las lentes se evitó el uso de gran angular, empleándose la mayor distancia focal provista por el *zoom* óptico de la cámara. Para lograr que las imágenes fueran nítidas a pesar del movimiento de la pesa, se empleó el mínimo tiempo de exposición posible. Para ello se dispuso una iluminación adecuada, producto de la combinación de tubos fluorescentes y luz solar.

## 3. Análisis de las fuerzas involucradas

Las fuerzas que actúan sobre la pesa mientras oscila son la magnética, la fuerza elástica, una fuerza disipativa y el peso, siendo la fuerza neta que actúa sobre la pesa:

$$\vec{F}_{neta} = \vec{F}_B + \vec{F}_E + \vec{F}_D + M\vec{g} \quad (1)$$

Donde $\mathbf{F_B}$ es la fuerza magnética, $\mathbf{F_E}$ la fuerza elástica, $\mathbf{F_D}$ la fuerza disipativa y $M\mathbf{g}$ el peso.

Para modelar el comportamiento del sistema se partió de las siguientes hipótesis:
a) El movimiento de rotación de la pesa es despreciable.
b) La fuerza magnética que actúa sobre la pesa depende únicamente de su posición.
c) El resorte tiene un comportamiento lineal y su masa es mucho menor que la de la pesa.
d) La fuerza disipativa depende linealmente de la velocidad de la pesa y se asocia al rozamiento, siendo la principal fuente de disipación de energía.

---
[6] CANON PS-SD750 (lente: 17,4 mm, resolución: 640 × 480, FPS: 30 cuadros por segundo).

Para resolver la ecuación de movimiento de la pesa, primero se determinaron las características de cada una de las fuerzas que actúan sobre la misma.

### 3.1 Estudio de la fuerza magnética

Para determinar la fuerza magnética que actúa sobre la pesa en función de su posición, se empleó el montaje de la figura 3. Primero se colocaron los imanes de barra junto al disco sobre un recipiente de poliestireno expandido, y el conjunto sobre una balanza digital. A continuación, se taró la misma y luego se colgó la pesa de un hilo. Posteriormente se modificó gradualmente la distancia entre la pesa y los imanes de barra, registrándose para las distintas distancias los valores indicados por la balanza. Dados estos valores —que son consecuencia del cambio de la normal sobre los imanes— se determinó la fuerza magnética (aplicando el principio de interacción se deduce que el módulo de esta última es igual al de la variación de la normal).[7] [5]

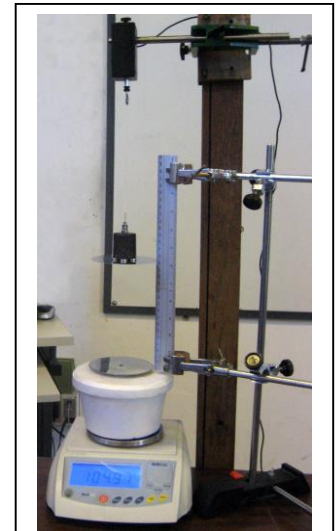

3. Montaje experimental para determinar la fuerza magnética sobre la pesa.

La gráfica de la figura 4 muestra los resultados experimentales de la fuerza magnética actuante sobre la pesa ($F_B$) en función de la distancia a los imanes de barra ($y$). Se observó que los valores experimentales se ajustan satisfactoriamente a una función potencial:

$$F_B(y) = C\, y^n \quad (2)$$

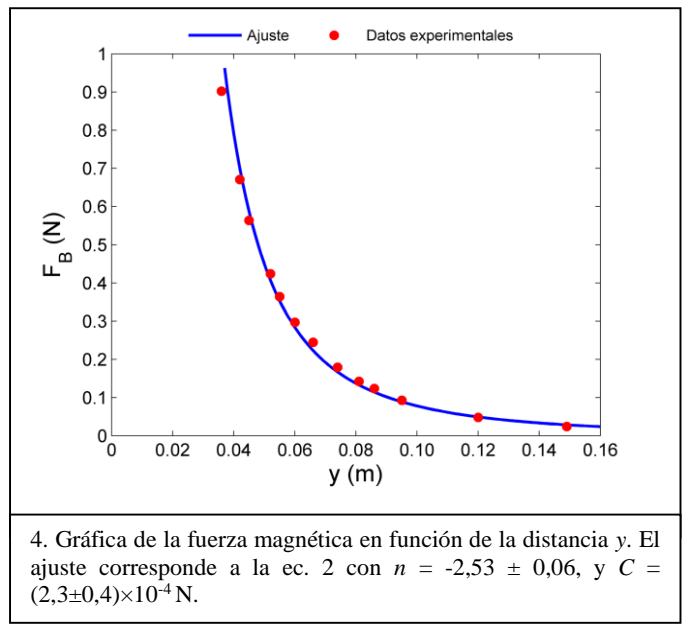

4. Gráfica de la fuerza magnética en función de la distancia $y$. El ajuste corresponde a la ec. 2 con $n = -2{,}53 \pm 0{,}06$, y $C = (2{,}3 \pm 0{,}4) \times 10^{-4}$ N.

Como se puede apreciar en la figura 5, $y(t)$ puede expresarse en función de $z(t)$:

$$y(t) = L - h - z(t) \quad (3)$$

---

[7] La exposición de la balanza a campos magnéticos muy intensos podría llegar a alterar las medidas, por lo que para aumentar la distancia entre los imanes de barra y el plato de la balanza se colocó entre ambos un recipiente de poliestireno expandido.

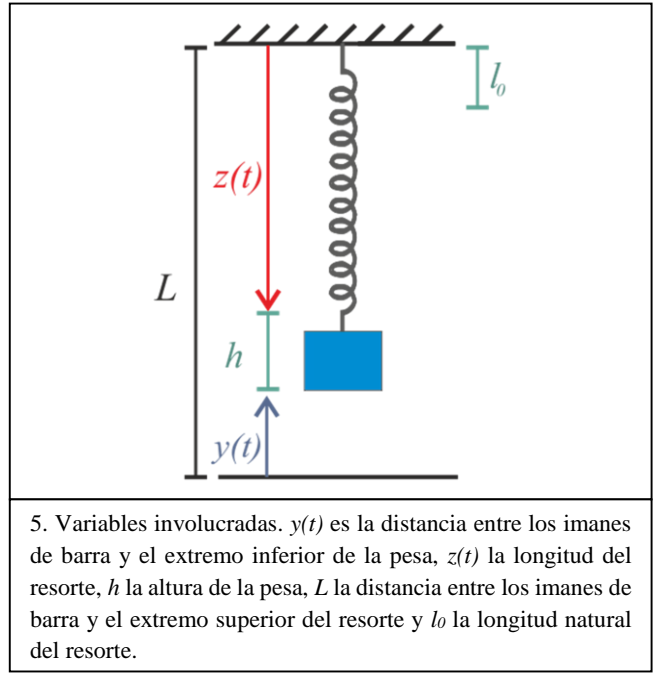

5. Variables involucradas. *y(t)* es la distancia entre los imanes de barra y el extremo inferior de la pesa, *z(t)* la longitud del resorte, *h* la altura de la pesa, *L* la distancia entre los imanes de barra y el extremo superior del resorte y $l_0$ la longitud natural del resorte.

Tras realizar este cambio de variable en la ecuación 2, la expresión de la fuerza magnética queda dada por:

$$F_B(t) = C(L - h - z(t))^n \quad (4)$$

**3.2 Estudio de la fuerza elástica y del amortiguamiento**

Para determinar las características de la fuerza restauradora y del rozamiento se realizó un montaje similar al de la figura 1 pero sin los imanes de la base. En estas condiciones las únicas fuerzas que actúan sobre la pesa son la ejercida por el resorte, el peso y la fricción del aire:

$$\vec{F}_{neta} = \vec{F}_E + \vec{F}_D + M\vec{g} \quad (5)$$

Suponiendo que el resorte es lineal y que la fuerza disipativa es proporcional a la velocidad de la pesa y tomando el origen de coordenadas en la posición de equilibrio de la misma, la ecuación 5 toma la forma:

$$M\ddot{x} = -kx - b\dot{x} \quad (6)$$

Donde *x* es la posición de la pesa respecto al punto de equilibrio, *b* el coeficiente de amortiguamiento, *k* la constante del resorte y *M* la masa del sistema conformado por la pesa junto al disco y los 5 imanes cilíndricos.

La ley horaria correspondiente a la ecuación 6 es:

$$x(t) = Ae^{-\beta t}sen(\omega t + \delta) \quad (7)$$

Donde:

$$\beta = \frac{b}{2M} \quad (8)$$

$$\omega = \sqrt{\left(\frac{k}{M}\right)^2 - \beta^2} \quad (9)$$

En la gráfica de la figura 6 se puede observar el ajuste de los datos experimentales a la ecuación 7 con los siguientes parámetros:

$$A = (1,365 \pm 0,001) \times 10^{-2} \text{ m}$$
$$\beta = (5,73 \pm 0,01) \times 10^{-3} \text{ s}^{-1}$$
$$\omega = (6,264 \pm 0,001) \text{ s}^{-1}$$
$$\delta = 1,4758 \pm 0,0001$$

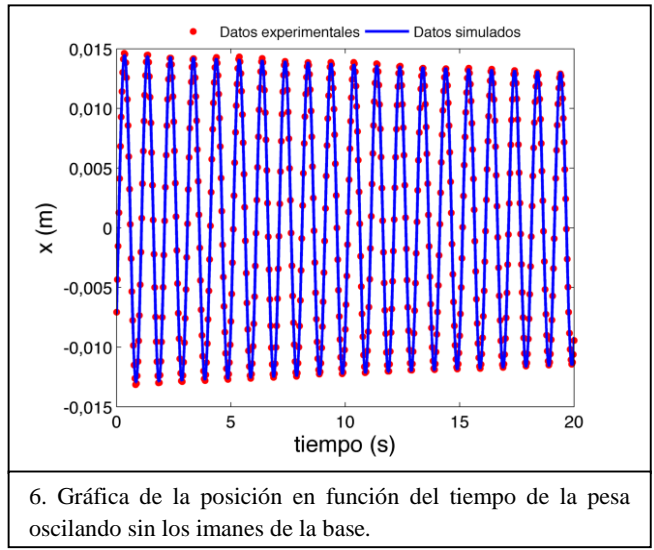

6. Gráfica de la posición en función del tiempo de la pesa oscilando sin los imanes de la base.

A partir de la ecuación 9 se podría determinar la constante elástica del resorte ($k$) si su masa fuera nula.[8] Como en el sistema estudiado la masa del resorte no es despreciable, la constante elástica se aproxima a la siguiente expresión [6]:

$$k = \left(M + \frac{m_{resorte}}{3}\right)\left(\omega^2 + \beta^2\right) \quad (10)$$

Sea $z(t)$ la posición del extremo inferior del resorte (ver figura 5). En función de esta coordenada la fuerza elástica está dada por:

$$F_E(t) = -k(z(t) - l_0) \quad (11)$$

Donde $l_0$ es la longitud natural del resorte:

$$l_0 = (7,1 \pm 0,2) \times 10^{-2} \text{ m}$$

A partir de los valores de la masa del sistema oscilante ($M = [257,9 \pm 0,1] \times 10^{-3}$ kg) y de la masa del resorte ($m_{resorte} = [22,76 \pm 0,01] \times 10^{-3}$ kg), la constante elástica resulta:

$$k = (10,42 \pm 0,03) \text{ N/m}$$

---

[8] La constante elástica también pudo haberse determinado por un método estático. Se optó por este procedimiento, ya que a partir del mismo se podía hallar directamente la constante elástica y el coeficiente de amortiguamiento.

La fuerza disipativa está dada por la siguiente ecuación:

$$F_D(t) = -b\,\dot{z}(t) \quad (12)$$

Donde $b$ es el coeficiente de amortiguamiento. De la ecuación 8 se desprende que:

$$b = 2M\beta$$
$$b = (2{,}96 \pm 0{,}06)\times 10^{-3}\,\text{kg/s}$$

### 3.3. Estudio dinámico del sistema

La fuerza neta que actúa sobre la pesa está dada por la ecuación 1:

$$\vec{F}_{neta} = \vec{F}_B + \vec{F}_E + \vec{F}_D + M\vec{g}$$

Combinando la ecuación anterior con 4, 11 y 12:

$$F_{neta} = C(L-h-z)^n - k(z-l_0) - b\dot{z} + Mg \quad (13)$$

Aplicando la 2ª Ley de Newton se arriba a la ecuación del movimiento:

$$\ddot{z} = \frac{C(L-h-z)^n - k(z-l_0) - b\dot{z}}{M} + g \quad (14)$$

La cual no es lineal y no posee solución analítica.

## 4. Análisis numérico y resultados experimentales de la posición y de la aceleración en función del tiempo

### 4.1 Análisis numérico, método de Euler

La ecuación diferencial del movimiento de la pesa no es lineal y no posee solución analítica. Para obtener la aceleración y la posición de la pesa en distintos instantes de tiempo, se resolvió numéricamente la ecuación de movimiento del sistema (14) utilizando el método de Euler en una planilla de cálculo Excel[9]. Este método numérico de resolución de una ecuación diferencial es el más simple de todos. Dado el intervalo de tiempo total para el cual se va a estudiar el movimiento, se divide este en $N$ intervalos $\Delta t$, tal que $\Delta t = t_{total}/N$. A dichos intervalos se los denomina "pasos". Para el cálculo de la velocidad, se considera constante la aceleración en cada "paso". Posteriormente, se considera constante la velocidad para determinar la posición. A partir de dichas suposiciones y del conocimiento de las condiciones iniciales del movimiento es posible calcular la velocidad y posición para distintos instantes de tiempo con el siguiente algoritmo:
1. Para un instante de tiempo $t_0$ y las condiciones iniciales $x_0$ y $v_0$, se determina la aceleración $a_0$ del objeto a partir de la ecuación de movimiento.

---

[9] Con el *software* Tracker también se puede resolver numéricamente la ecuación diferencial del movimiento utilizando el "modelo dinámico de la partícula", el cual emplea el método de Runge-Kutta. Dicha opción no fue utilizada, ya que uno de los objetivos de este trabajo es mostrar la potencialidad y sencillez del método de Euler.

2. Dada $a_0$ y definido el "paso de tiempo" $\Delta t$, se calcula la velocidad que tendría el objeto en un tiempo $t_0+\Delta t$, la cual asumiendo una aceleración constante $a_0$ durante todo el intervalo de tiempo $\Delta t$, resulta: $v_1(t_0+\Delta t)=v_0+a_0\Delta t$.
3. Conocida la velocidad $v_1$, se determina la posición $x_1$, tal que $x_1(t_0+\Delta t)=x_0+v_1\Delta t$.
4. Se realiza el proceso iterativo hasta completar el tiempo total del análisis. [2]

El método de Euler no es el más eficiente para resolver una ecuación diferencial. Sin embargo, se puede mejorar parcialmente su precisión reduciendo el valor del paso entre cada operación. Esto no se puede hacer en forma ilimitada, ya que la reducción del paso implica una mayor cantidad de cálculos y un aumento del error introducido por el redondeo. De todas formas, la riqueza del método radica en su sencillez y fácil compresión por parte de los estudiantes, así como en la posibilidad de obtener resultados razonablemente buenos para intervalos de tiempo no muy grandes.

### 4.2 Análisis de la posición, la velocidad y la aceleración en función del tiempo

Se estudió la evolución del sistema en torno a la posición de equilibrio estable. Las gráficas de las figuras 7 y 8 permiten evidenciar las diferencias notables entre el sistema estudiado y el de un oscilador armónico. En todas las gráficas se superpusieron los datos experimentales y los valores simulados mediante cálculo numérico. El paso utilizado para la simulación fue de 0,005s.

Como se puede apreciar en la figura 7, los valores experimentales de la posición se corresponden muy satisfactoriamente con los simulados. En cuanto a las características del movimiento, al trabajar experimentalmente dentro de la región de equilibrio estable, naturalmente el movimiento del sistema es oscilatorio. Con la particularidad de que no es simétrico en torno a la posición de equilibrio, como se observa en el gráfico de la figura 7. Del mismo gráfico también se desprende que el tiempo que tarda la pesa en ir desde la posición más cercana a los imanes hasta la posición de equilibrio es mayor que el que tarda en ir desde la posición de equilibrio a la posición más alejada de los imanes. Asimismo, la distancia entre la posición de la pesa más cercana a los imanes y la posición de equilibrio es mayor que la distancia entre la posición de equilibrio y la posición de la pesa más alejada de los imanes. Respecto al período de oscilación del sistema, este es mayor que el que posee el sistema en ausencia de la fuerza magnética.

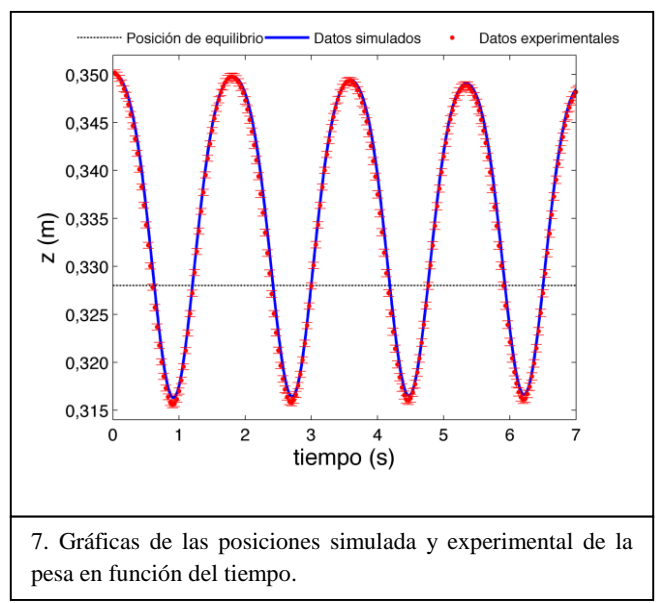

7. Gráficas de las posiciones simulada y experimental de la pesa en función del tiempo.

En la figura 8 se muestran los gráficos de las velocidades simulada y experimental de la pesa en función del tiempo (arriba) y de las aceleraciones simuladas y experimental en función del tiempo (abajo). De los mismos se evidencia la diferencia notable con un movimiento armónico simple. En lo que concierne al gráfico de la aceleración obtenido de la simulación, este muestra una notable diferencia con el del oscilador armónico, presentando en un ciclo cuatro cambios de concavidad. Esto es consecuencia de la posición en la que se colocaron los imanes, así como del hecho de que la fuerza magnética depende de la posición.

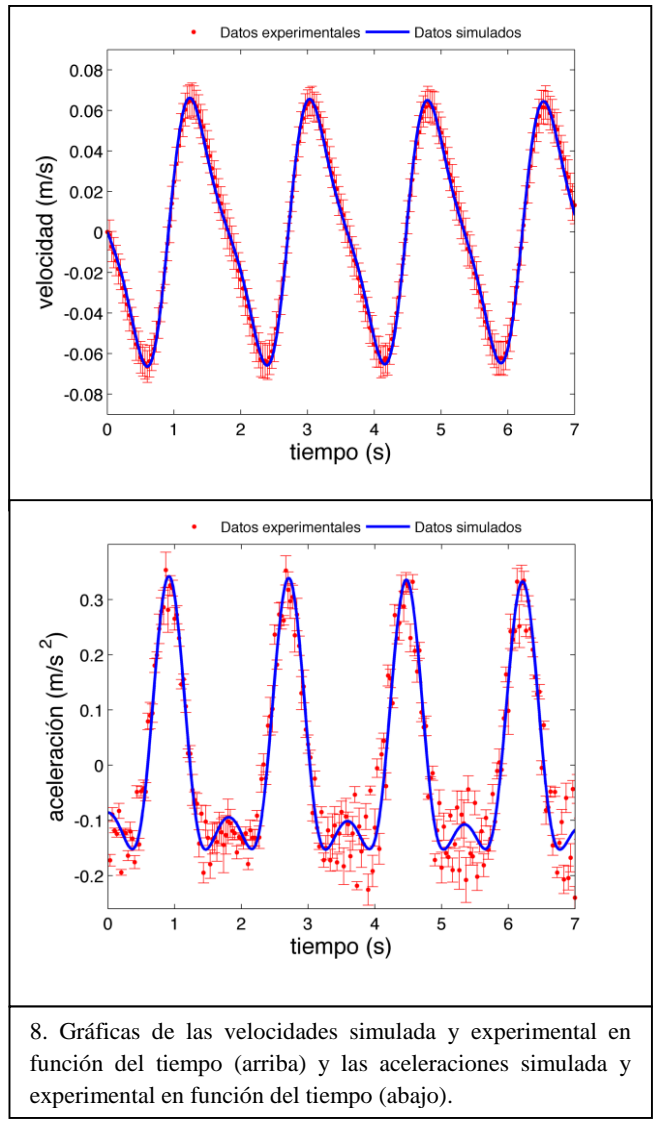

8. Gráficas de las velocidades simulada y experimental en función del tiempo (arriba) y las aceleraciones simulada y experimental en función del tiempo (abajo).

La gran incertidumbre asociada a la determinación de la aceleración mediante análisis de video no permite una comparación certera de los valores experimentales con los simulados. Una posible mejora del dispositivo experimental consistiría en incorporarle un acelerómetro. Esto sería factible empleando un *smartphone* como parte del sistema y registrando los valores arrojados por sus sensores de aceleración [7,8].

## 5. Conclusiones y perspectivas

El oscilador fue estudiado teórica y experimentalmente desde el punto de vista dinámico. Al comparar dichas predicciones con los datos experimentales, la correspondencia fue satisfactoria. Esto permite inferir que el sistema analizado puede ser abordado teóricamente, resolviendo su ecuación de movimiento con el método de Euler haciendo uso de una planilla de cálculo.

El experimento presentado permite explorar un fenómeno que no puede ser descrito adecuadamente con ecuaciones resolubles analíticamente. Este tipo de sistema generalmente no es abordado en los cursos básicos de Física. Su estudio ayuda a que los alumnos comprendan que no todos los problemas poseen una solución exacta. Al mismo tiempo, los acerca al uso de métodos numéricos para la resolución de ecuaciones diferenciales. Asimismo, el sistema analizado es de fácil construcción y permite explorar características de sistemas oscilatorios no analizadas usualmente como la dependencia de la frecuencia de oscilación con la amplitud del movimiento.

La utilización de una planilla de cálculo facilita la implementación del algoritmo de Euler por parte de los estudiantes, aunque no posean conocimientos avanzados de programación. Por otra parte, les permite comparar de forma sencilla las predicciones del modelo con los datos experimentales y observar qué ocurre al variar manualmente cada uno de los parámetros.

Finalmente, el trabajo realizado genera un abanico de posibilidades para desarrollar y profundizar a futuro, entre las cuales se destacan: el estudio de los espacios de fase del sistema, el análisis de oscilaciones forzadas, y la utilización de un acelerómetro para el estudio experimental del movimiento.